\documentclass[manuscript,nonacm]{acmart}

\newcommand{\perception}{\textit{Perception}}
\newcommand{\moderation}{\textit{Moderation}}
\newcommand{\negotiation}{\textit{Negotiation}}
\usepackage{graphicx} 
\usepackage[utf8]{inputenc}
\usepackage{kotex}
\usepackage{enumitem} 
\usepackage{subcaption}
\usepackage{multirow}
\usepackage{amsmath} 
\usepackage{booktabs} 
\usepackage{threeparttable} 
\usepackage{array} 
\usepackage{circledsteps}
\usepackage{verbatim} 
\usepackage{ulem}

\newif\ifdebug
\newif\ifdiff
\newif\iffinaldiff

\debugtrue

\newcommand{\sysname}{\texttt{Chatperone}}

  {\list{}{\leftmargin=#1\rightmargin=#1}\item[]}%
  {\endlist}

\definecolor{bluebox_fill}{cmyk}{0.27, 0.01, 0.01, 0}
\definecolor{bluebox_border}{cmyk}{0.79, 0.44, 0.02, 0.01}

\newcounter{unknowncounter}
\newcounter{commentcounter}
\newcounter{needchartcounter}
\newcounter{needfigcounter}
\newcounter{needtablecounter}
\newcounter{needcitecounter}

\newcommand{\drop}[1]{\ifdiff{\color{pink}{}}\else{}\fi}

\AtBeginDocument{%
  }

\title{Chatperone: An LLM-Based Negotiable Scaffolding System for Mediating Adolescent Mobile Interactions}
\author{Suwon Yoon}
\email{suwon.yoon@postech.ac.kr}
\orcid{0009-0002-6665-1626}
\affiliation{%
  \department{Department of Computer Science and Engineering}
  \institution{POSTECH}
  \streetaddress{Cheongam-ro 77}
  \city{Pohang}
  \state{Gyeongbuk}
  \country{South Korea}
  \postcode{37673}
}

\author{Seungwon Yang}
\email{sw.yang@postech.ac.kr}
\orcid{0009-0009-0755-2450}
\affiliation{%
  \department{Department of Computer Science and Engineering}
  \institution{POSTECH}
  \streetaddress{Cheongam-ro 77}
  \city{Pohang}
  \state{Gyeongbuk}
  \country{South Korea}
  \postcode{37673}
}

\author{Jeongwon Choi}
\email{choijw@postech.ac.kr}
\orcid{0009-0000-7849-2365}
\affiliation{%
  \department{Department of Computer Science and Engineering}
  \institution{POSTECH}
  \streetaddress{Cheongam-ro 77}
  \city{Pohang}
  \state{Gyeongbuk}
  \country{South Korea}
  \postcode{37673}
}

\author{Wonjeong Park}
\email{iamwj@postech.ac.kr}
\orcid{0000-0003-1457-0859}
\affiliation{%
  \department{Center for Mobile Embedded Software Technology}
  \institution{POSTECH}
  \streetaddress{Cheongam-ro 77}
  \city{Pohang}
  \state{Gyeongbuk}
  \country{South Korea}
  \postcode{37673}
}

\author{Inseok Hwang}
\email{i.hwang@postech.ac.kr}
\orcid{0000-0001-7370-3944}
\affiliation{%
  \department{Department of Computer Science and Engineering}
  \department{Center for Mobile Embedded Software Technology}
  \institution{POSTECH}
  \streetaddress{Cheongam-ro 77}
  \city{Pohang}
  \state{Gyeongbuk}
  \country{South Korea}
  \postcode{37673}
}

\begin{abstract}
    Adolescents' uncontrolled exposure to digital content can negatively impact their development. Traditional regulatory methods, such as time limits or app restrictions, often take a rigid approach, ignoring adolescents' decision-making abilities. Another issue is the lack of content and services tailored for adolescents. To address this, we propose \sysname{}, a concept of a system that provides adaptive scaffolding to support adolescents. \sysname{} fosters healthy mobile interactions through three key modules: \perception{}, \negotiation{}, and \moderation{}. This paper outlines these modules' functionalities and discusses considerations for real-world implementation.
\end{abstract}


\begin{teaserfigure}
    \centering
    \includegraphics[width=0.88\textwidth]{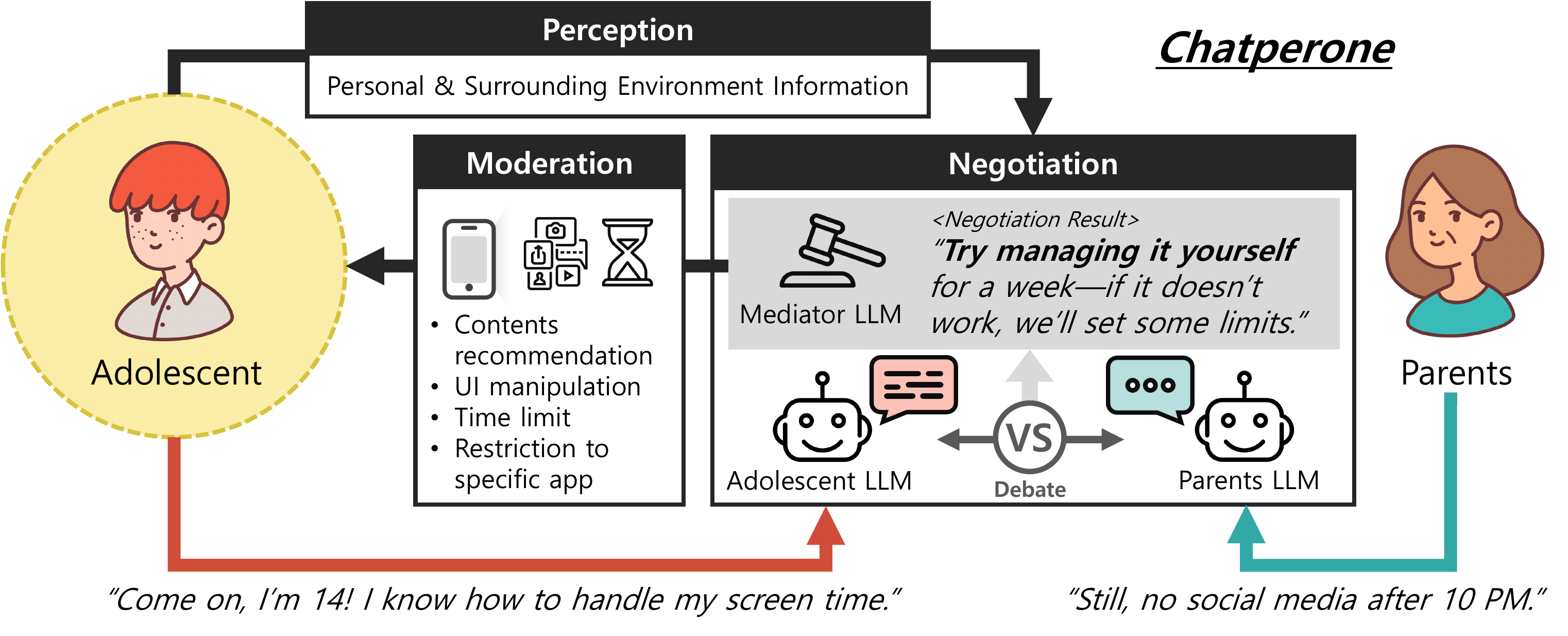}
    \vspace{-10pt}
    \caption{Interactions between adolescents and parents facilitated by \sysname{}: \perception{}, \negotiation{}, \moderation{}.}
    \label{fig:teaser}
\end{teaserfigure}

\begin{document}

\maketitle

\section{Introduction}
\label{sec:introduction}

The recent increase in smartphone usage among teens has raised significant social concerns not only regarding their exposure to inappropriate or unfiltered content but also concerning addiction~\cite{care2011clinical, holloway2013zero, lanette2018much}. The traditional approach to regulating teens' smartphone use involves restrictions, either in the form of parental controls that limit screen time or legal age requirements for social media registration~\cite{wang2021protection, o2011impact}. However, such rigid and one-size-fits-all regulatory approaches often fail to consider the emotional and social contexts of individual adolescents.

Services such as YouTube Kids implement age-based filtering, applying various levels of restrictions for each age group to ensure a safe and age-appropriate viewing experience for children~\cite{alqahtani2023children}. This approach corresponds to Lev Vygotsky’s cognitive development theory, particularly the concept of the Zone of Proximal Development (ZPD) which emphasizes the scaffolding needed to enhance learners’ abilities effectively~\cite{shabani2010vygotsky}.
However, as adolescents move beyond the pre-teen stage, they often begin using the same online platforms and services as adults, often with minimal guidance. Therefore, similar to the approach used for children, adolescents also require scaffolding that supports their interests and zone of proximal development.

At the same time, adolescents need appropriate scaffolding while maintaining autonomy in setting their own goals.
This aligns with Erikson’s psychosocial development theory, which characterizes adolescence as a stage of identity versus role confusion~\cite{erikson1968identity}. 
During this period, teenagers examine their values, roles, and personal beliefs, fostering the development of a cohesive sense of self~\cite{balleys2020searching}. Successfully navigating this developmental stage depends on the ability to make and learn from significant choices, which fosters a healthy identity formation.

In this regard, it is essential not only to guide adolescents but also to grant them an appropriate level of autonomy over the content they consume. Allowing adolescents to negotiate or self-regulate the scope and nature of their online activities fosters a sense of ownership and responsibility, both vital for identity development. Parents should provide scaffolding instead of imposing rigid rules---by setting reasonable boundaries and fostering continuous dialogue, feedback, and intervention when necessary~\cite{wisniewski2017parental, vaala2015monitoring}.

We anticipate that combining mobile technology and large language models (LLMs) can play a pivotal role. With the pervasiveness of mobile ubiquitous technology, continuous and real-time sensing of users’ physiological signals, movements, and contextual data has become increasingly feasible~\cite{seneviratne2017survey, mohr2017personal}.
Leveraging this stream of real-time sensing data, LLMs can enhance their contextual understanding, enabling more adaptive and intelligent interactions. LLMs exhibit remarkable proficiency in natural interactions via text and speech, as well as synthesizing information to perform complex reasoning tasks~\cite{zhang2024can, patil2025advancing, chai2025text, ju2025emosync}. For example, MindShift~\cite{wu2024mindshift} aimed to provide interventions for problematic smartphone use based on LLMs, considering the user’s current context such as events, emotions, and personal goals.

In this paper, we propose \sysname{}---a concept for an LLM-based negotiable scaffolding mobile system for adolescents. Our key idea is to put an LLM-based mediator between adolescents and parents, enabling a nuanced consideration of both perspectives and contextual factors. As shown in Figure~\ref{fig:teaser}, we propose three main modules to ensure adaptive scaffolding: \perception{}, \negotiation{}, and \moderation{}. \sysname{} leverages \perception{} to assess the adolescent's current state and surrounding environment. 
By integrating sensing data and the perspectives of individual stakeholders, \sysname{} conducts \negotiation{} via LLM. The resulting decision is then conveyed to adolescents as a suitable form of \moderation{}.
We hope that our approach could inspire further discussions on technologies that support healthy mobile interactions while respecting adolescents' decision-making autonomy. 
The following sections outline the main components and potential applications of \sysname{} (\S\ref{sec:system}), along with considerations for its real-world implementation (\S\ref{sec:discussion}).

\section{Negotiable Scaffolding System}
\label{sec:system}

\subsection{System Overview}

This paper envisions the conceptual architecture of \sysname{}, a flexibly configurable solution encompassing content recommendation/moderation and UI manipulation in all aspects of smartphone use.
\sysname{}'s scaffolding features consists of three main modules: \perception{}, \negotiation{}, and \moderation{}.

The \perception{} module gathers information on what the adolescent is exposed to by utilizing various smartphone functionalities (e.g., microphone, social media apps). The goal is to understand what the adolescent is already familiar with, what they have not yet been exposed to, and the broader cultural and social contexts surrounding their media consumption. Similar approaches have been explored in past HCI research~\cite{lee2024open}. Due to security and privacy concerns, this module operates locally on the device.

The \negotiation{} module facilitates scaffolding by serving as an intermediary between adolescents and parents through an LLM-powered system. This module enables both parties to engage in natural language conversations and negotiations with the system to dynamically adjust the degree of moderation as needed. 
It is aligned with the Teen Online Safety Strategies Framework (TOSS) in two aspects: active mediation and self-regulation~\cite{wisniewski2017parental}.
Since parents cannot always respond to adolescents' requests, the system offers extensive active mediation opportunities through an LLM that represents the parents. 
For self-regulation, the adolescent can understand the inner reasoning process of a mediator LLM and be motivated to become more responsible.

The \moderation{} module enforces actual constraints on the content adolescents see and the actions they take. Based on interaction logs observed by the \perception{} module, the \negotiation{} module determines whether moderation is necessary. For example, if the \perception{} module detects that the adolescent has not yet been exposed to violent or offensive content, the \moderation{} module blocks related media and replaces it with more appropriate alternatives. 



\subsection{Usage Scenario}

Leo has been spending late nights on his phone, often watching videos. Concerned about his sleep schedule, his mother turns to {\sysname} to set a reasonable smartphone curfew. Before enforcing the restriction, {\sysname} analyzes Leo’s past behavior. It acknowledges his recent late-night usage has been excessive. Based on this, {\sysname} supports the mother’s request and implements a curfew that gradually limits access after a certain hour. Leo, frustrated by the restriction, argues that he can manage his own bedtime. Rather than outright denying him, {\sysname} offers a structured compromise---if he can demonstrate better nighttime habits over the next week, the curfew can be adjusted. Meanwhile, the system allows access to essential functions, such as emergency calls, ensuring that restrictions do not feel overly punitive.

Consider another scenario. Alex has spent hours studying, though his parent is unaware. Feeling he’s earned a break, he decides to spend the night chatting with his friend. Seeing him on his phone late at night, his parent asks {\sysname} to reduce his screen time. Before enforcing the request, {\sysname} analyzes Alex’s recent behavior. It recognizes his prior study efforts and balanced usage patterns. Instead of immediately restricting access, the system rules against the parent’s request, explaining that based on Alex’s past behavior, limiting his social time would be unjustified. To maintain fairness, {\sysname} suggests a compromise: allowing extended screen time but with gradual nudges to encourage winding down for the night.


\section{Discussion}
\label{sec:discussion}

\subsection{Incorporating Teens’ Voices in System Design}

When implementing this system, it is very important to incorporate the teenager's perspective in the system design. While the system might seem reasonable and balanced from a parental or developer standpoint, teenagers may perceive it as restrictive, unfair, or intrusive if their needs and autonomy are not fully respected.

\subsection{Privacy Considerations}

A key concern with pervasive systems is that they not only record data about the direct user but also capture interactions involving other individuals. This raises ethical and practical concerns about data privacy, agency, and unintended surveillance. One potential way to mitigate such concerns is through abstraction via LLMs. If the recorded data is processed in a way that abstracts or generalizes specific details, some privacy risks could be alleviated. 
Another possible concern arises when a pervasive system captures multimodal data (e.g., text, audio, video, sensor inputs) extensively. The abundance of data makes indexing and retrieving the necessary information a significant challenge. Determining what information is essential and under what circumstances it should be accessed remains an open problem.

Our current approach envisions preprocessing through multiple AI agents to manage and filter recorded data before storage or retrieval. However, we are uncertain whether this is the optimal strategy. Further investigation is needed to determine whether agent-based preprocessing effectively balances privacy, efficiency, and usability.

\subsection{Ensuring Fairness in Negotiation with LLMs}

One major challenge with the LLM-mediated negotiation system is the LLM's susceptibility to external influence and manipulation. LLMs can be easily swayed by the framing of an argument, user prompts, or even adversarial inputs~\cite{raina2024llm}, raising concerns about whether they can actually provide effective mediation. Although there have been attempts to use AI for structured debates and negotiation~\cite{abdelnabi2023llm}, these approaches still face limitations.

Our vision is to reframe negotiation processes in a way that limits the direct manipulation of decision-making agents. Instead of allowing users to interact with the final judgment system, LLMs should serve as intermediaries. By structuring LLMs as negotiation proxies rather than decision-makers, we may be able to reduce their vulnerability to manipulation and create a more robust negotiation process. However, further research is needed to explore how different models handle persuasion, bias, and external pressure in multi-agent negotiation settings.

Furthermore, the system could collect perspectives and comments from people around the adolescent to gain a broader understanding of the situation.
The impact of this engagement should be evaluated through further research.




\section{Conclusion}
\label{sec:conclusion}

This paper reimagines how a mobile system can mediate digital autonomy and family dynamics in teenage smartphone use. Instead of rigid parental controls, it envisions a conceptual system where teenagers negotiate content exposure with AI, fostering a more personalized and developmentally appropriate approach. Rather than imposing restrictions, the system acts as a scaffold, gradually adjusting content access based on readiness, ensuring that self-regulation is cultivated rather than enforced.
Beyond individual autonomy, this system positions the LLM as a mediator in parent-teen conflicts over mobile use. By reframing control as negotiation, \sysname{} aims to transform tension into collaboration, reducing friction and fostering mutual understanding. At the same time, it acts as a protective buffer, regulating exposure to inappropriate content without resorting to blanket prohibitions.
This is an invitation to rethink digital guardianship as an evolving process that adapts alongside its users. Balancing autonomy and protection remains a challenge, yet this work marks a step toward a more thoughtful and human-centered approach to online content moderation.
\begin{acks}
This work was supported by the National Research Foundation of Korea(NRF) grants funded by the Korea government(MSIT) (RS-2025-00553946, RS-2024-00451947).
This research was also partly supported by Seoul R\&BD Program(SP240008) through the Seoul Business Agency(SBA) funded by The Seoul Metropolitan Government, and a grant (RS-2025-00564342) from the Korea Institute for Advancement of Technology (KAIT), funded by the Ministry of Trade, Industry and Energy (MOTIE), Republic of Korea.
\end{acks}

\bibliographystyle{ACM-Reference-Format}
\bibliography{references/default, references/smartphone, references/llm}

\end{document}